\documentclass[amsmath,twocolumn]{aastex702}

\shorttitle{Active Learning for Molecular Discovery}
\shortauthors{R.\,I.\,El-Kholy \& Z.\,M.\,Hayman}

\submitjournal{ApJ}

\usepackage{tabularx}

\begin{document}

\title{Active Learning for Molecular Discovery: A Retrospective Test toward TMC-1}

\correspondingauthor{R.\,I.\,El-Kholy}
\email{relkholy@sci.cu.edu.eg, reham.elkholy@cu.edu.eg}

\author[orcid=0000-0001-6337-1384]{R.\,I.\,El-Kholy}
\affiliation{Department of Astronomy, Space Science, and Meteorology, Faculty of Science, Cairo University, 1 Gamaa Street, Giza 12613, Egypt}
\email{relkholy@sci.cu.edu.eg, reham.elkholy@cu.edu.eg}  

\author[orcid=0009-0009-4770-5626]{Z.\,M.\,Hayman} 
\affiliation{Department of Astronomy, Space Science, and Meteorology, Faculty of Science, Cairo University, 1 Gamaa Street, Giza 12613, Egypt}
\email{zmhayman@cu.edu.eg}

\begin{abstract}
Machine-learning recommendation lists can help prioritize searches for new interstellar molecules, but whether re-ranking such a list as detections accumulate improves on it has not been tested. We present a pre-registered retrospective test toward TMC-1, using a published machine-learning list of 1,510 recommended candidates, consolidated to 1,434 candidates, and the species reported there through 2026 September. At four annual origins from 2022 to 2025, a logistic-regression ranking on seven released features, refit on the detections known at each origin, separates later detections from never-detected candidates with a pooled AUC of 0.867, against 0.706 for the best single-criterion baseline (difference 0.160; 95\% interval 0.095--0.230). The advantage persists under a shifted date boundary and two alternative feature regimes, but is not established among the top 100 candidates (12 versus 7 detections). A searcher that refits on its own observations reaches half of the 63 detected species on the list in a median of 200 observations, against 332 for the best fixed ordering and 717.5 expected for random search. Uncertainty sampling, the standard active-learning rule, performs worse than random (AUC 0.210): later detections lie at low predictive uncertainty, close to known chemistry. Of 101 species in our validation set, 37 are absent from the list, and only 13 of these appear in a pool rebuilt from its source databases, including 2 of 11 radicals and neither ion. Candidate generation, not ranking, is the main constraint. We release 100 recommended targets, with code and frozen inputs, for prospective testing.
\end{abstract}


\section{Introduction} \label{sec:intro}
The Taurus Molecular Cloud 1 (TMC-1) is among the most chemically rich regions of space known~\citep{Lee2021}, and its inventory continues to grow quickly through surveys such as GOTHAM on the Green Bank Telescope~\citep{McGuire2020} and QUIJOTE on the Yebes 40\,m telescope~\citep[e.g.][]{Cernicharo2022}. Newly detected species span charged molecules~\citep{Cernicharo2022,Silva2023,Remijan2023} and large polycyclic aromatic hydrocarbon derivatives~\citep{Wenzel2025}. Deciding which molecules to search for next is a bottleneck in its own right: as the space of chemically plausible candidates grows combinatorially with molecular size, chemical intuition alone does not scale, and candidates must instead be identified systematically~\citep{Lee2021}.

\citet{Lee2021} introduced a machine-learning approach to this problem (the \texttt{umda} pipeline\footnote{\url{https://github.com/laserkelvin/umda}}): molecules are embedded in a continuous vector space with mol2vec~\citep{Jaeger2018}, candidates are selected by proximity to species already detected, and regressors, including a Gaussian process, predict their column densities. The approach has since been extended to other sources and embeddings~\citep{Fried2023,Scolati2023,ToruShay2025}, coupled to observational detectability~\cite{ToruShay2025a} and automated line assignment~\citep{Fried2026}, and applied to other targets such as IRC+10216~\citep{Li2026}. Across this line of work, a candidate list is built once from similarity to known chemistry and then left static.

\citet{ToruShay2025} provided the retrospective check most directly relevant here: of the 53 non-isotopologue TMC-1 species detected after \citeauthor{Lee2021}'s training cutoff, the original 2021 recommendations contained 13 (24.5\%) and retraining the updated pipeline on the same pre-2021 data raised this to 24 (45.3\%).\footnote{Matching \citeauthor{ToruShay2025}'s Table~C1 detections against \citeauthor{Lee2021}'s released list by molecular structure~(Section~\ref{subsec:candidate}) places substantially more than 13 of the 53 in the list; we do not attempt to reconcile the two counts.} This establishes recovery of later detections as a meaningful benchmark, but it is a tally of list membership for two fixed pipelines. It does not address how a list should be ordered, or whether a system should update as new detections arrive rather than staying fixed.

This is the setting active learning is designed for: a model is fit on the labels available so far, used to prioritize what to observe next, and refit as new labels arrive. It has not yet been applied this way in this line of work; scoring has instead come from a single similarity search~\citep{Lee2021} or a single trained regressor~\citep{ToruShay2025}, computed once. A natural score for such prioritization is a model's predictive confidence, available in these pipelines through the posterior variance of a Gaussian process or the disagreement among an ensemble, but it has so far only been used to discard uncertain predictions~\citep{Fried2023} or reported without being used to select targets~\citep{ToruShay2025}.

In this paper, we test, retrospectively, whether an actively updated ranking finds later TMC-1 detections faster than the static orderings this line of work has used. We use the embeddings, pre-cutoff training data and per-candidate predictions released with the \texttt{umda} pipeline~\citep{Lee2021}, and evaluate candidates drawn from \citeauthor{Lee2021}'s own released recommendation list after removing entries that duplicate the pre-cutoff training set. Using a series of temporal cutoffs, we compare a ranking that combines predictive confidence with structural and physical features against the similarity and abundance orderings available at each cutoff, and we test an online variant that updates from its own revealed outcomes, in the manner of an active-learning loop. We report both what this comparison shows and what it does not, including the practical size of any gain over the simplest orderings an observer would already use. Section~\ref{sec:methods} describes the model, candidate universe, and evaluation design; Section~\ref{sec:results} reports the results, including a list of recommended next targets; Section~\ref{sec:disc} discusses what the comparison establishes and its limitations; Section~\ref{sec:conc} concludes. The code and data underlying this analysis are available on GitHub\footnote{\url{https://github.com/rehamelkholy/TMC-1}} and archived at Zenodo~\citep{zenodo}.

\section{Data and Methods} \label{sec:methods}
\subsection{Pre-cutoff Model} \label{subsec:precut}
The model we evaluate against is the \texttt{umda} pipeline of \citet{Lee2021}. Molecules are represented by 300-dimensional mol2vec vectors~\citep{Jaeger2018}, trained on 3,316,454 unique structures and compressed to 70 dimensions by principal component analysis. The model is trained on the column densities of the 87 species known toward TMC-1 at the time of that work (their Table~A2). The inventory is expanded by bootstrap resampling with Gaussian noise ($\sigma = 0.5$) added to the logarithmic column densities, to 800 observations. Several regressors are fit to these data. They include a Gaussian process regressor (GPR) with a rational-quadratic, dot-product and white-noise kernel. \citeauthor{Lee2021} selected 1,510 candidate molecules as the nearest neighbors of the known species in the embedding~(Section~\ref{subsec:candidate}). For each candidate they released the nearest known species and its distance in the embedding, the GPR's predicted log column density and its $1\sigma$ uncertainty, and quantum-chemical properties: dipole moment, molecular weight, charge and spin multiplicity \citep{Lee2021a}. We use the second version of this release~(2021 July 28); an earlier version~(2021 July 7) lists the same candidates with different predicted column densities and uncertainties. For 148 candidates the geometry optimization did not converge~\citep{Lee2021}; we use the released values as given.

These released values are the features used throughout this work. They are what an observer following this pipeline would have had in hand after 2021. We therefore take them from the release rather than regenerating them. To validate the pipeline, we reconstructed it from the public repository and compared its predictions with \citeauthor{Lee2021a}'s released values for the 13 later-detected molecules tabulated by \citet{ToruShay2025}; all 13 agree to within 0.005~dex. The predicted column densities printed in their Table~4 also agree with the release for 12 of the 13. For HC$_7$NH$^+$, the printed value~($9.18 \times 10^{11}$~cm$^{-2}$) equals the released prediction for C$_7$N rather than that for HC$_7$NH$^+$~($1.2 \times 10^{11}$~cm$^{-2}$).

For the sensitivity analysis of~Section~\ref{subsec:robust}, which recomputes these features as new detections accumulate, we refit the GPR on the training observations stored with the distributed model, optimizing its kernel hyperparameters by maximum likelihood; the distributed model's hyperparameters are at their initial values. At the 2021 state, the refit reproduces \citeauthor{Lee2021a}'s released uncertainties and nearest-species distances closely (Spearman $\rho=0.98$) and its predicted column densities less closely ($\rho=0.79$), as expected when hyperparameters are re-optimized. Section~\ref{subsec:robust} reports results with the refit features, both held at the 2021 state and updated as detections accumulate.

\subsection{Validation Set} \label{subsec:valid}
The validation set comprises molecules reported toward TMC-1 that are absent from the 87-species training inventory of \citet{Lee2021}. A species is included if it is reported toward TMC-1, is not an isotopologue, is a firm rather than tentative detection, and has a SMILES that encodes the connectivity, hydrogen placement and net charge given in its discovery paper. We began from Table~C1 of \citet{ToruShay2025}, which lists detections of species absent from that inventory through 2023 September. We identified all other reports through 2026 September 21 by searching the NASA Astrophysics Data System (ADS) and screening every returned title and abstract.\footnote{The queries were \texttt{abs:("TMC-1" OR "TMC1" OR "Taurus Molecular Cloud 1") pubdate:[2021-08 TO 2026-09] collection:astronomy}, returning 239 records, and \texttt{author:("Cernicharo, J" OR "McGuire, B" OR "Agundez, M") pubdate:[2021-08 TO 2026-09] collection:astronomy}, returning 252 records, both run on 2026 September 24, and \texttt{abs:("TMC-1" OR "TMC1" OR "Taurus Molecular Cloud 1") pubdate:[2013-01 TO 2021-07] collection:astronomy}, returning 169 records, run on 2026 September 30.} To find earlier detections absent from the training inventory, we also checked every species attributed to TMC-1 in the census of interstellar molecules of \citet{McGuire2022}, the TMC-1 line survey of \citet{Kaifu2004} and the compilation of \citet{Agundez2013}; searched ADS for TMC-1 reports from 2013 January to 2021 July; matched the species inventory of a recent review of TMC-1 chemistry~\citep{Agundez2026}; and searched ADS individually for every remaining candidate with three or fewer heavy atoms. Our title-and-abstract screening misses detections reported only in the body of papers about other species; the review's inventory identified three such cases (C$_7$H, NCO and CH$_3$SH), all included.

Applying the criteria to Table~C1 removed three species. Two are structurally identical to members of the training inventory (HCCCO and HC$_3$O; c-C$_9$H$_8$ and C$_9$H$_8$), and one was reported toward G+0.693 rather than TMC-1 (C$_4$H$_8$). The search added three detections within the Table~C1 period that the table omits: HSO, CH$_3$NC and ethynylbenzene. We also corrected SMILES that misrepresented the reported species' structure or charge; for example, H$_2$CCCN had been encoded as vinyl cyanide. Reports of l-H$_2$C$_3$O, which is in the training inventory, and of cis-HCOOH, which has the same SMILES as the training species HCOOH, were not included.

The final set contains 101 species~(Table~\ref{tab:validation}). Each is assigned a first-public date: the arXiv posting date where one exists, otherwise the journal acceptance date, otherwise the online publication date. Month-level precision is sufficient for the annual cutoffs used in~Section~\ref{subsec:design}. Forty-four species were first reported before the first cutoff, 2022 January 1, 38 of them before \citet{Lee2021} was published; all 44 are treated as known at every cutoff.

\begin{table*}
	\caption{Molecules reported toward TMC-1 that are absent from the training inventory of \citet{Lee2021}.}
	\label{tab:validation}
	\setlength{\tabcolsep}{4pt}
	\begin{tabularx}{\textwidth}{lccccclX}
		\hline\hline
		Species	& SMILES & Date	& Date source & Lee ID	& Rule	& Reference	& Note\\
		\hline
		CCH	& [C]\#C	& 1980-08-01	& online	& 503	& A	& \citet{Wootten1980} &	Pre-2021\\
		H$_2$S	& S	& 1989-10	& online	& 307	& A	& \citet{Minh1989}	& Pre-2021\\
		HCNH$^+$	& [CH+]=N	& 1991-07	& online	& 290	& A	& \citet{Schilke1991} &	Pre-2021; Earliest TMC-1 report located\\
		NO &	[N]=O	& 1993-01	& online	& 275	& A & \citet{Gerin1993}	& Pre-2021; Earliest TMC-1 report located\\
		NS &	[N]=S	& 1994-01	& online	& 279	& A	& \citet{McGonagle1994} &	Pre-2021\\
		CO	& [C-]\#[O+]	& 1997-09-10	& online	& 800	& A	& \citet{Pratap1997} &	Pre-2021; Earliest TMC-1 report located\\
		\hline
	\end{tabularx}
	\tablecomments{Table~\ref{tab:validation} is published in its entirety in the machine-readable format. A portion is shown here for guidance regarding its form and content. Date is the first-public date; Lee ID is the entry number in the recommendation list of \citet{Lee2021a};  Rule gives the matching rule of~Section~\ref{subsec:candidate} (A, identical SMILES; B, identical after removing stereochemistry; C, same formula and hydrogen-resolved connectivity); ``Pre-2021" refers to detections reported before \citet{Lee2021} was published.}
\end{table*}

\subsection{Candidate Universe} \label{subsec:candidate}
The candidates are the 1,510 molecules recommended by \citet{Lee2021a}, selected as the union of the 100 nearest neighbors in the embedding of each known species, after removing duplicates, species containing elements other than H, C, N, O, Si, P and S, and van der Waals complexes. Restricting the universe to this list means the comparison concerns how to order the candidates this pipeline actually produced, not how a candidate list should be generated.

Some entries in the list are structurally identical to species already in the training inventory, written with different SMILES strings. We compiled the 87 training species from Table~A2 of \citet{Lee2021}, cross-checked against Table~B1 of \citet{ToruShay2025}, and removed every list entry matching one of them by canonical SMILES, InChIKey or the connectivity rule given below; this removed 27 entries, five of them written in charge-separated forms that only the connectivity rule identifies, and one a neutral HCCN entry that the table's SMILES, written with a spurious positive charge, had concealed. The list also represents some single species by more than one entry, for example as different resonance forms. We merged entries sharing a species key, defined as the molecular formula including charge together with the connectivity and hydrogen layers of the InChI. For merged entries, numeric features are averaged, and a candidate counts as charged or open-shell if any of its entries is. Seven entries whose SMILES fail standard valence checks (hypervalent protonated cations) are kept as separate candidates. The result is 1,434 candidates.

A species in the validation set is matched to a candidate if their canonical SMILES are identical; or if they are identical after removing stereochemistry, provided the list entry carries no stereochemical marks; or if they share the molecular formula, including charge, and the heavy-atom connectivity with the same hydrogen count on each atom, again only where the list entry carries no stereochemical marks. The last rule allows for bond-order and charge-placement differences between Lewis structures of the same species. Under these rules, 64 of the 101 species match a candidate. The matched list entries and the rule applied are given in~Table~\ref{tab:validation}. The cis and trans isomers of CH$_3$CHCHCN match the same stereo-agnostic entry, so the 64 species correspond to 63 positive candidates. The remaining 37 species do not appear in \citeauthor{Lee2021}'s list, so no ordering of the list can place them among the top candidates; they are excluded from the evaluation, and we return to this limitation in~Section~\ref{subsec:limitations}.

\subsection{Temporal Design} \label{subsec:design}
We evaluate each scoring method as if it had been used at a series of past dates. The cutoffs, or origins, are the first days of 2022, 2023, 2024 and 2025. They were chosen from detection counts alone, before any model was fit, under the pre-specified rule that the first origin is dropped if fewer than 12 positive candidates are known before it. At an origin, a candidate counts as known if the species it matches has a first-public date before the origin; species reported before the first origin, including those reported before \citet{Lee2021} was published~(Section~\ref{subsec:valid}), are therefore known at every origin. The test window of each origin runs to the next origin, and the last window ends on 2026 September 21. The test pool at an origin consists of all candidates not known at that origin. Its test positives are those whose matched species was first reported inside the window. Candidates detected after the window remain in the pool as negatives, as they would have appeared to an observer at that date. At the four origins, 34, 44, 50 and 59 candidates are known. The test pools contain $1{,}400$, $1{,}390$, $1{,}384$ and $1{,}375$ candidates, of which 10, 6, 9 and 4 are test positives, respectively, 29 in total.

\subsection{Scoring Methods} \label{subsec:scoring}
The combined ranking is an L2-regularized logistic regression ($C=1.0$) with seven features: the distance to the nearest known species, the GPR uncertainty, the GPR predicted log column density, the dipole moment, the molecular weight, and binary indicators for charged and open-shell species. The features are standardized over all candidates. At each origin the model is refit with known candidates labeled 1 and all others 0, so the ranking is updated as detections accumulate. Candidates are ordered by the fitted score. The model is fit with scikit-learn~\citep{Pedregosa2011}, with no hyperparameter tuning.

We compare it with three baselines, fixed in advance as representing orderings an observer would plausibly use. The first orders candidates by predicted column density. The second orders them by a detectability proxy, the predicted log column density plus twice the logarithm of the dipole moment, reflecting that rotational line strength scales with column density and the square of the dipole moment; candidates with zero dipole moment are ranked last. The third orders candidates by similarity, with the smallest distance to a known species first. The best baseline for each comparison is the one with the most detections among its first 100 candidates, pooled over origins. Two uncertainty-based rules are reported for comparison but do not enter the evaluation criteria: uncertainty sampling, which ranks the highest GPR uncertainty first, and an upper-confidence-bound rule, which ranks by predicted log column density plus one standard deviation. A random ordering provides the chance reference; its expected performance is computed exactly. Where tied scores straddle a cutoff, counts are the exact expectation under random tie-breaking.

\subsection{Feedback Simulation} \label{subsec:feed-meth}
The temporal design refits the ranking once per origin. To test learning from the outcomes of one's own observations, the setting active learning is designed for, we also simulate a searcher that reveals candidates in batches and learns only from what it has revealed. The pool is all 1,434 candidates, and the 63 positive candidates count as detections regardless of date. The searcher starts with no labels. It reveals 5 candidates per round, the first two rounds at random. After each round it refits a class-balanced, L2-regularized logistic regression ($C=0.3$) on the same standardized features, using all outcomes revealed so far, and reveals the 5 highest-scoring unrevealed candidates next. Each slot is replaced by a random unrevealed candidate with probability 0.1, and batches are drawn at random until at least one positive has been revealed. We record the number of observations needed to reveal 32 positives, half of the total, over 200 random seeds, and compare it with the expected number for each baseline ordering and for random selection.

\subsection{Evaluation Criteria} \label{subsec:eval}
The analysis plan, including the evaluation criteria below, was written down and time-stamped before any evaluation was run; later clarifications were recorded as dated amendments. Both are included in the data release. Three criteria were specified in advance. First (V1, ranking quality), the combined ranking's AUC, pooled across origins as the mean of the per-origin values weighted by test positives, exceeds that of the best baseline. Second (V2, practical yield), at a budget of 100 candidates per origin the combined ranking finds at least 20\% more test positives, pooled over origins, than the best baseline. Third (V3, active learning), the feedback searcher's median number of observations to reach half of the detections is lower than that of the best baseline ordering.

For V1 and V2, we estimate uncertainty with 2,000 paired bootstrap resamples. Within each origin, positives and negatives are resampled separately with replacement, and the same resample is used for both methods. A criterion passes if it is met and the 95\% interval on the difference excludes zero; it is {\sc inconclusive} if the point estimate favors the combined ranking but the interval includes zero; and it fails otherwise. With about 25 pooled test positives, the number available when the plan was written, a simulation-based power analysis gives roughly 93\% power for V1 if the true AUCs were 0.90 against 0.75, and roughly 52\% if they were 0.85 against 0.75. We also specified two sensitivity analyses in advance: one removes test positives whose date does not come from an arXiv posting, and the other moves every test positive dated within 60 days of an origin to the other side of it.

Several changes were made after the first results had been seen; none altered the criteria, the models or the baselines. When the two independent implementations described below were first compared, the definition of the species key proved to admit two readings; we fixed it in writing and the verdicts were identical under both. A second pass matching list entries to the training inventory by InChIKey removed 12 entries the first pass had missed, and checks for charge errors and charge-separated representations removed six more; these are included in the 27 of~Section~\ref{subsec:candidate}. The literature search of Section~\ref{subsec:valid} added four species (CH$_3$NC, HS$_2$, ethynylbenzene and CH$_3$CO$^+$); the census, survey, compilation and review checks added seventeen species reported before the first origin (C$_5$N, l-HC$_4$N, NS$^+$, HC$_3$S$^+$, CO, HCO$^+$, HCO, CCH, H$_2$S, NO, NS, HCNH$^+$, HNCS, HC(O)CN, OH, HSCN and HCS) and three test positives reported after it (C$_7$H, NCO and CH$_3$SH); and a subsequent check that each reference reports the first firm TMC-1 detection moved three dates earlier (CH$_2$CCH, HCOOCH$_3$ and CH$_2$CCHC$_3$N), none of them across an origin boundary for any candidate in the list. All analyses were re-run on the final universe and validation set, and every number reported here comes from that final version. V2's outcome is sensitive to these corrections: in intermediate versions of the validation set its interval on the difference excluded zero, and in the final version it does not. After the final results were fixed, we replaced the simulated random tie-breaking by its exact expectation; no verdict changed, and no reported value changed by more than 0.01.

The analysis plan also specified a sensitivity analysis with the GPR features recomputed at each origin from all species then known. As specified, it included each known species among its own reference species and GPR training data, giving every training positive a zero distance that no test candidate can have; we therefore replace it with a post hoc leave-self-out version, and add a post hoc version using the refit GPR held at the 2021 state, which separates the effect of the refit from the effect of updating. Results of the original specification are available in the data release.

All reported results were computed by two independent implementations of the analysis plan, and neither set of results was examined until both were complete. The two agree exactly for every reported quantity, except for the first date sensitivity, for which one implementation approximated the rule identifying arXiv dates; there we report the values from the exact implementation. All reported results were computed in a single software environment; with a different version of the optimization library they change by less than 0.001 in AUC and by at most one detection within the top 100.

\section{Results} \label{sec:results}
\subsection{Ranking Quality} \label{subsec:ranking}
\begin{figure}
	\centering
	\includegraphics[width=\columnwidth]{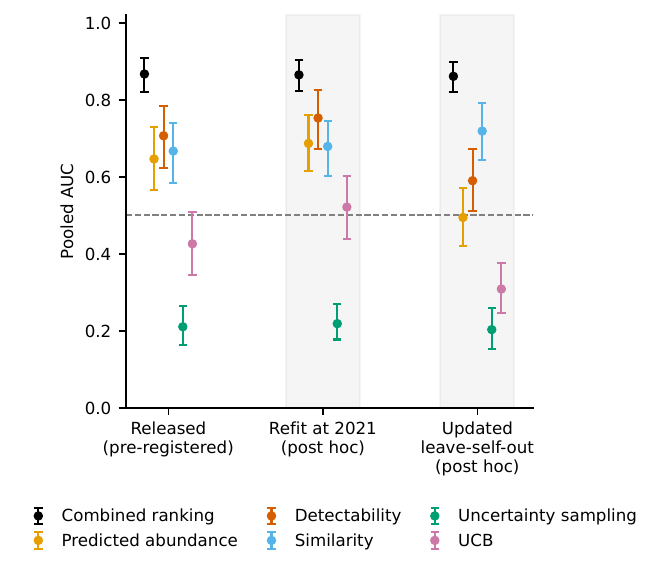}
	\caption{Pooled AUC of each ordering, with 95\% bootstrap intervals, for three ways of computing the features: the values released by \citet{Lee2021}, used in the pre-registered analysis; our refit GPR held at its 2021 state (post hoc); and features recomputed at each origin from all species then known, excluding each candidate's own species (post hoc). The dashed line marks chance.}
	\label{fig:1}
\end{figure}

Figure~\ref{fig:1} shows the pooled AUC of each ordering. With the released features, the pre-registered analysis, the combined ranking reaches 0.867 (95\% interval 0.819--0.909). The best baseline, the detectability proxy, reaches 0.706 (0.623--0.783), and predicted abundance and similarity reach 0.646 and 0.667, respectively. The combined ranking's advantage over detectability is 0.16 (95\% interval 0.10--0.23), so criterion V1 passes. The advantage holds at the first three origins, where the combined ranking's AUC is 0.83, 0.87 and 0.91 against 0.69, 0.65 and 0.64 for detectability. It reverses at the last origin, which has four test positives: there detectability reaches 0.97 against 0.87 for the combined ranking~(Section~\ref{subsec:yield}).

The two uncertainty-based rules rank later detections worse than random selection. Uncertainty sampling reaches 0.210 (0.163--0.264), placing the eventual detections systematically near the bottom of the list, and the upper-confidence-bound rule reaches 0.426 (0.344--0.508). The molecules later detected toward TMC-1 are thus those about which the model is most, not least, confident; we return to this in Section~\ref{subsec:uncertain}.

\subsection{Practical Yield} \label{subsec:yield}
\begin{figure}
	\centering
	\includegraphics[width=\columnwidth]{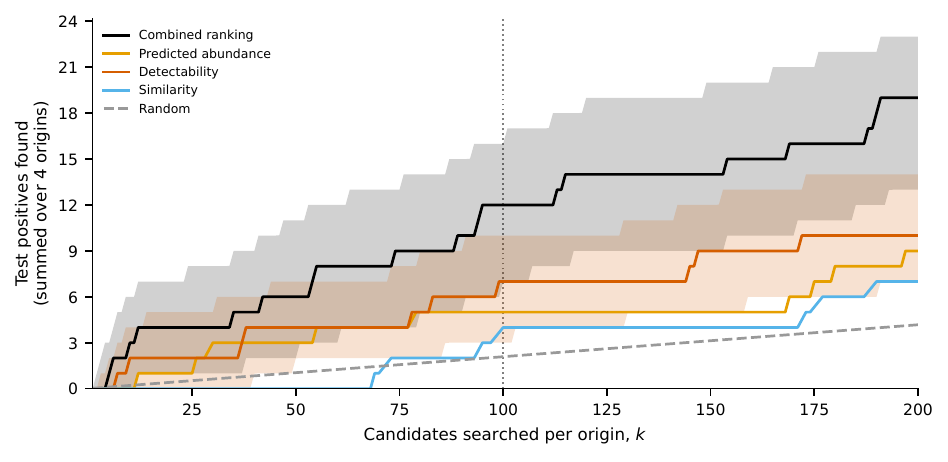}
	\caption{Test positives found among the first $k$ candidates at each origin, summed over the four origins (29 test positives in total), for the released features. Shaded bands are 95\% bootstrap intervals for the combined ranking and for detectability, the best baseline. The dashed line is the expected yield of random selection, and the dotted line marks the pre-registered budget of 100 candidates.}
	\label{fig:2}
\end{figure}

Figure~\ref{fig:2} shows how many test positives each ordering finds as a function of the number of candidates searched per origin. With 100 candidates per origin, the combined ranking finds 12 of the 29 test positives, against 7 for detectability, 5 for predicted abundance, 4 for similarity, and an expected 2.1 for random selection. The gain over detectability, a factor of 1.7, exceeds the 20\% threshold for V2. However, the 95\% interval on the difference, 0 to 10, includes zero, so V2 is {\sc inconclusive} as specified~(Section~\ref{subsec:eval}).

Table~\ref{tab:ranks} gives the rank each test positive received at its origin. The combined ranking places half of them within its first 154 candidates, against 327 for detectability and 403 for similarity, and it ranks 24 of the 29 higher than detectability does. Its advantage is concentrated at the first three origins, where it finds 10 of 25 test positives within 100 candidates, against 3 for detectability. At the last origin the pattern reverses: detectability ranks all four test positives within the first 100 candidates, three of them (CH$_3$CHS, HS$_2$ and CH$_2$CHCHS) within its first 40, while the combined ranking places two within 100.

\begin{table}
	\caption{Rank of each test positive in the test pool of its origin, under the combined ranking and two baselines (released features).}
	\label{tab:ranks}
	\setlength{\tabcolsep}{4pt}
	\centering
	\begin{tabular}{lclll}
	\hline\hline
	Species	& Origin &	Comb. &	Detect. &	Simi.\\
	\hline
	cis-/trans-CH$_3$CHCHCN &	2022 &	12 &	83 &	692\\
	NC$_4$NH$^+$ &	2022 &	74 &	315 &	641\\
	C$_7$H &	2022 &	113 &	616 &	1205\\
	c-C$_5$H$_4$CCH$_2$ &	2022 &	154 &	516 &	1032\\
	HCCNCH$^+$ &	2022 &	191 &	145 &	436.5\\
	C$_3$H$^+$ &	2022 &	224 &	342 &	172.5\\
	CH$_2$CHCH$_2$CN &	2022 &	267 &	848 &	313\\
	C$_5$H$^+$ &	2022 &	303 &	221 &	251.5\\
	HC$_7$NH$^+$ &	2022 &	361 &	486 &	286\\
	CH$_2$C(CH$_3$)CN &	2022 &	748 &	785 &	692\\
	CH$_3$NC &	2023 &	5 &	260 &	466.5\\
	C$_{10}$H$^-$ &	2023 &	10 &	249 &	489\\
	CH$_3$CHCO &	2023 &	89 &	481 &	403.5\\
	H$_2$CCCH$^+$ &	2023 &	115 &	502 &	176\\
	E-1-C$_4$H$_5$CN &	2023 &	274 &	327 &	1068.5\\
	C$_2$H$_5$CHO &	2023 &	591 &	1073 &	285\\
	HCNS &	2024 &	6 &	10 &	349.5\\
	HC$_5$N$^+$ &	2024 &	42 &	147 &	622.5\\
	HC$_3$N$^+$ &	2024 &	54 &	78 &	99\\
	CH$_2$(CN)$_2$ &	2024 &	55 &	759 &	972.5\\
	HC$_7$N$^+$ &	2024 &	94 &	172 &	622.5\\
	(Z)-NCCHCHCN &	2024 &	169 &	1334 &	815.5\\
	C$_5$H$_4$ &	2024 &	188 &	551 &	307.5\\
	NCO &	2024 &	190 &	885 &	69\\
	CH$_3$CH$_2$CCH &	2024 &	398 &	539 &	276.5\\
	CH$_3$CHS &	2025 &	35 &	7 &	189\\
	CH$_3$SH &	2025 &	95 &	99 &	94.5\\
	HS$_2$ &	2025 &	295 &	37 &	72\\
	CH$_2$CHCHS &	2025 &	325 &	38 &	403\\
	\hline
	\end{tabular}
	\begin{flushleft}
		\tablecomments{Origins are 1 January of the year given; the test pools contain $1{,}400$, $1{,}390$, $1{,}384$ and $1{,}375$ candidates. Tied candidates receive their mean rank. The cis and trans isomers of CH$_3$CHCHCN match a single list entry.}
	\end{flushleft}
\end{table}

\subsection{Feedback Simulation} \label{subsec:feed-res}
\begin{figure}
	\centering
	\includegraphics[width=\columnwidth]{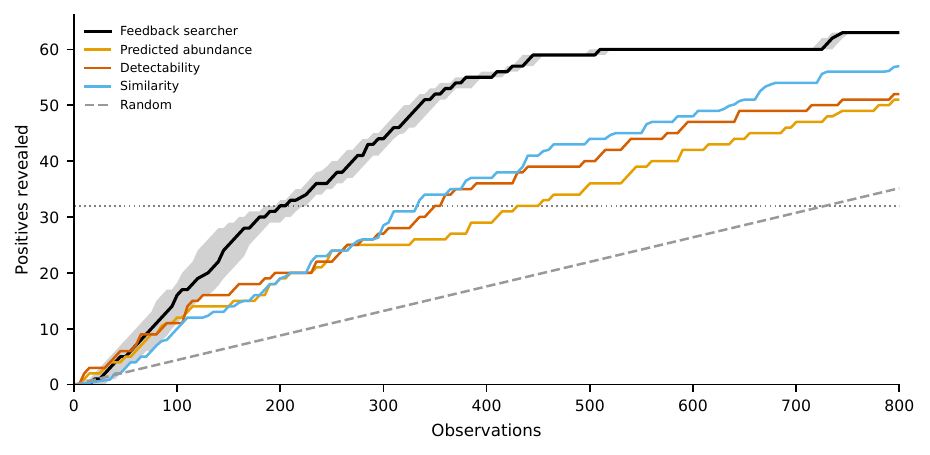}
	\caption{Detections revealed as a function of the number of observations, for released features. The feedback searcher refits its ranking on its own outcomes every five observations; its line is the median over 200 random seeds and the band the interquartile range. Fixed orderings are averaged over exact expectations over tied orderings, and the dashed line is the expected yield of random selection. The dotted line marks half of the 63 detections. The random reference shows the expected number of detections recovered at each budget; the expected number of observations needed to reach 32 is 717.5.}
	\label{fig:3}
\end{figure}

Figure~\ref{fig:3} shows the number of detections revealed by the feedback searcher as observations accumulate. The searcher reveals half of the 63 detections after a median of 200 observations (interquartile range 185--221 over 200 random seeds). The best fixed ordering, similarity, needs 332 on average; detectability and predicted abundance need 348 and 428, and random selection 717.5. The searcher is faster than every fixed ordering in all 200 seeds, so criterion V3 passes. Starting without labels, it reveals 16 detections in its first 100 observations, against 11 for detectability. By 300 observations it has revealed 44, against 27, and it reveals all 63 within 745 observations.

\subsection{Robustness} \label{subsec:robust}
The two date sensitivities leave the comparison unchanged. Moving every test positive dated within 60 days of an origin to the other side of it gives pooled AUCs of 0.874 for the combined ranking and 0.701 for detectability, with 11 and 7 test positives found within 100 candidates, respectively. Removing test positives whose date does not come from an arXiv posting gives 0.862 and 0.691, with 10 and 5, respectively. As specified, we report these as point estimates.

Figure~\ref{fig:1} also shows the two post hoc feature regimes. With the refit GPR held at its 2021 state, the combined ranking reaches 0.865 (0.822--0.902) and detectability 0.752 (0.762--0.825), so the gain narrows to 0.11 (0.04--0.18). At 100 candidates the two find the same number of test positives, 10. With features recomputed at each origin, the combined ranking reaches 0.861 (0.819--0.898). Among the baselines, detectability falls to 0.590 (0.511--0.673) and similarity becomes the strongest at 0.718 (0.644--0.790). The gain over similarity is 0.14 (0.07--0.22), and within 100 candidates the combined ranking finds 10 test positives against 4, although the 95\% interval on that difference ($-1$ to 11) includes zero. The ranking advantage is therefore established in all three regimes, and the practical-yield advantage in none. In both post hoc regimes uncertainty sampling again ranks detections below chance, at 0.218 and 0.203.

Across the three regimes, the combined ranking's AUC varies by less than 0.01, from 0.861 to 0.867, while that of the best baseline ranges from 0.71 to 0.75 depending on how the features are computed.

\subsection{Recommended Targets} \label{subsec:recommend}
To provide a list of targets for future searches, we refit the combined ranking on all 63 positive candidates known through 2026 September 21, using the released features, and ranked the remaining $1{,}371$ candidates. Table~\ref{tab:targets} lists the highest-ranked candidates. The full list of the top 100 is available online, with each candidate's predicted column density, dipole moment, the features contributing most to its rank, and any detection in other sources. Two candidates whose SMILES do not describe physical structures (a triple bond within a three-membered ring, and five cumulated double bonds within a five-membered ring) were removed, leaving gaps in the ranks.

\begin{table*}
	\caption{Highest-ranked candidates for detection toward TMC-1.}
	\label{tab:targets}
	\centering
	\begin{tabularx}{\textwidth}{cllllcXX}
	\hline\hline
	Rank &	Species &	Class &	$\log{N}$ &	$\mu(D)$ &	Catalog &	Detected elsewhere &	Note\\
	\hline
	1 &	C$_9$H &	radical &	9.88 &	15.54 &	CDMS\,109501 &	 &  \\	
	2 &	C$_5$N$^+$ &	ion &	12.00 &	11.10 &	 &	 & \\	
	3 &	HCNCC &	closed-shell &	12.05 &	9.40 &	 &	 &	uncommon isomer\\
	4 &	HSiN &	closed-shell &	12.54 &	5.00 &	 &	 &	name corrected\\
	6 &	l-C$_7$H$_2$ &	closed-shell &	11.92 &	8.10 &	CDMS\,086501 &	 & \\	
	7 &	H$_2$CCNH &	closed-shell &	13.84 &	1.71 &	CDMS\,041503 &	Sgr B2(N) \citep{Lovas2006} & \\
	8 &	HONC &	closed-shell &	11.52 &	3.45 &	CDMS\,043512 &	 & \\	
	9 &	CH$_3$CC$^-$ &	ion &	12.23 &	7.00 &	 &	 & \\	
	10 &	l-C$_9$H$_2$ &	closed-shell &	11.67 &	9.80 &	 &	 & \\
	11 &	C$_3$N$^+$ &	ion &	12.62 &	7.20 &	 &	 & \\	
	\hline
	\end{tabularx}
	\begin{flushleft}
		\tablecomments{Table~\ref{tab:targets} is published in its entirety (100 candidates) in the machine-readable format; a portion is shown here for guidance regarding its form and content. $\log{N}$ is the column density (cm$^{-2}$) predicted by \citet{Lee2021}; $\mu$ is the dipole moment; Catalog gives the CDMS or JPL entry where laboratory rest frequencies exist. Ranks~5 and~71 were removed as non-physical structures. ``Name corrected" marks a species whose name differs from the formula label in \citeauthor{Lee2021}'s list.}
	\end{flushleft}
\end{table*}		

None of the 100 has been reported toward TMC-1. Twenty-five have been detected elsewhere, for example ketenimine (H$_2$CCNH, rank 7) and carbodiimide (HNCNH, rank 13) in Sgr~B2(N). These are natural first targets, since their emission has been seen in other sources. Thirteen of the 100, including C$_2$H$_2$, cannot be detected through rotational spectroscopy because they have no permanent dipole moment, and water (rank 12) is impractical to observe from the ground; we flag rather than remove them, because the ranking itself does not account for observability.

The main practical limitation lies elsewhere. Searching for a molecule requires laboratory rest frequencies, and only 23 of the top 100, and 4 of the top 10, have an entry in the CDMS~\citep{Mueller2005} or JPL~\citep{Pickett1998} spectroscopic catalogs. The highest-ranked candidates with catalog entries, and therefore searchable in existing survey data, are C$_9$H (rank 1), l-C$_7$H$_2$ (rank 6), H$_2$CCNH (rank 7), HONC (rank 8), HNCNH (rank 13) and H$_2$CCCHNC (rank 15). For most of the others, including the charged chains C$_5$N$^+$ and C$_3$N$^+$ near the top of the list, a search would first require laboratory spectroscopy.

The temporal tests of Sections~\ref{subsec:ranking}--\ref{subsec:robust} indicate how a ranking of this kind has performed on past detections; they do not predict how many of these candidates will be detected. The list covers only chemistry reachable from \citeauthor{Lee2021}'s candidate list~(Section~\ref{subsec:implications}), and it is archived in its present form so that future detections can test it.

\section{Discussion} \label{sec:disc}
\subsection{What This Establishes} \label{subsec:establish}
The pre-registered test asked whether a simple combined ranking, refit at each origin from the detections known at that time, orders \citeauthor{Lee2021}'s released recommendation list better than any single criterion would have. On the measure the protocol designated as primary, it does. Pooled over the four origins, the combined ranking reaches AUC $= 0.867$, against 0.706 for the strongest baseline (detectability). The difference is 0.160, with a 95\% bootstrap interval of 0.095 to 0.230. The result holds under the 60-day boundary shift (0.874 vs 0.701). It also holds in both post hoc feature regimes: features refit and held at their 2021 state (0.865 vs 0.752), and leave-self-out updated features (0.861 vs 0.718).

The more informative pattern is what changes across these regimes and what does not. The combined ranking's AUC moves by less than 0.01. The strongest baseline changes both its value and its identity: detectability in two regimes, similarity to known chemistry in the third. Under leave-self-out features, ranking by predicted abundance falls to AUC $=0.49$, no better than chance. The advantage of the combined ranking therefore does not depend on \citeauthor{Lee2021}'s abundance predictions being accurate for the species that were later detected. It comes from the combination of features, and Section~\ref{subsec:uncertain} suggests which of them carry it.

The advantage is not established for the top of the list. Within the 100 highest-ranked candidates, the combined ranking recovers 12 later detections against 7 for detectability, compared with a random expectation of 2.1. The interval on the difference (0 to 10) does not exclude zero, so V2 in {\sc inconclusive}. In the refit-2021 regime the two are tied at 10. As noted in Section~\ref{subsec:eval}, the V2 outcome also changed across intermediate label corrections. We therefore read the evidence as supporting a better overall ordering, not a demonstrated gain for a fixed observing budget of 100 targets. Both orderings recover detections at roughly three to six times the random rate in the top 100.

The feedback searcher (V3) shows that the same simple model, learning only from what it has itself ``observed", reaches half of the 63 detections in a median of 200 observations (interquartile range 185 to 221). The best fixed ordering, similarity, needs 332 observations, and random search would need 717.5 in expectation. All 200 seeds reach the target sooner than the best fixed ordering. This is the result most directly relevant to how a survey team would use such a tool: re-ranking as detections accumulate costs nothing beyond refitting a seven-feature logistic regression.

None of this shows that the model predicts which molecules are present in TMC-1. The outcome it learns is ``reported as detected by a given date". Section~\ref{subsec:limitations} discusses why that distinction matters.

\subsection{Why Uncertainty Sampling Fails} \label{subsec:uncertain}
Uncertainty sampling was included as a non-gating comparator because it is the default acquisition rule in active learning. It performs well below random, with a pooled AUC of 0.210, and it places no later detection within the top 200 at any origin. UCB, which adds uncertainty to predicted abundance, is intermediate (AUC $=0.426$). A post hoc, descriptive comparison of the 29 species detected from 2022 onward with the $1{,}371$ candidates never detected explains why. This analysis was not part of the pre-registered protocol and carries no inferential weight in the evaluation.

The GPR predictive standard deviation in \citeauthor{Lee2021}'s released feature is low for the later detections: median 2.08, against 3.70 for the never detected candidates (two-sided Mann-Whitney $p \approx 7 \times 10^{-8}$). The median later detection sits at the 21$^{\text{st}}$ percentile of the never-detected distribution. None of the 29 falls within the top tenth by uncertainty, and 22 fall within the bottom quarter. Uncertainty is strongly tied to how far a candidate lies from the training molecules and how large it is: across the full list, its Spearman correlation is 0.74 with embedding distance and 0.69 with molecular mass. The later detections are both closer to known chemistry (median distance 3.69 vs 4.77, $p\approx0.002$) and lighter (65 vs 96~amu, $p\approx1\times10^{-5}$) than the candidates never detected. Ranking by uncertainty therefore puts large, chemically unfamiliar molecules first, which are the ones least likely to be detected.

The combined model learns the opposite. Its standardized coefficient on uncertainty is the largest in magnitude at every origin, ranging from $-2.1$ to $-2.3$, and is stable from the 2022 origin to the final fit on all 63 detections. Reversing the uncertainty ranking would give a pooled AUC of about 0.79, below the combined ranking at every origin. That ordering is, moreover, available only in hindsight. The combined model arrives at the same direction prospectively: it assigns uncertainty a weight of $-2.2$ already at the 2022 origin, using only the 34 detections known at that time.

The median also suggests how the detected inventory grows. The detections known before 2022 lie closest to the training chemistry (median distance 2.74, mass 57~amu). The later detections lie further out, and the never-detected candidates furthest. Discovery toward TMC-1 has advanced outward from known chemistry in small steps, not through jumps to the most uncertain region of chemical space.

The failure is one of objective, not of the uncertainty estimate itself. Uncertainty sampling selects the points that would most improve a regression model. The task here is to select species that will be detected. In a campaign whose aim is to improve the abundance model, querying high-uncertainty candidates may still be the right choice. For target selection aimed at detection, our results indicate that predictive uncertainty is informative, but in the opposite direction from the one the standard acquisition rule assumes.

\subsection{Limitations} \label{subsec:limitations}
The test set is small. It contains 29 later detections across four windows (10, 6, 9 and 4), and the final window alone could not support any conclusion. The bootstrap intervals are correspondingly wide, and the sensitivity of V2 to individual label corrections~(Section~\ref{subsec:eval}) shows how much single species can move the early-retrieval measure.

The labels are positive and unlabeled, not positive and negative. A candidate not yet detected may be absent, below current sensitivity, or simply not yet searched for. The evaluation therefore measures how well a ranking anticipates reported detections, not how well it identifies what is present.

The outcome also reflects observational choices. Whether a species is reported depends on whether someone searched for it, and a search depends in turn on whether laboratory rest frequencies are available and whether its expected lines are bright enough. These conditions plausibly favor small, polar molecules close to known chemistry, which is the region the combined model rewards. \citeauthor{Lee2021}'s list has been public since 2021, and several of its authors are active in TMC-1 surveys. We have no evidence that the list influenced target choice, but we cannot exclude it. To the extent that it did, part of what the model learns is the search behavior of the community rather than the chemistry of the cloud.

Any re-ranking is limited by what the list contains. Of the 101 species in the validation set, 37 have no counterpart in \citeauthor{Lee2021}'s recommendations and could not be recovered by any ordering of it. \citeauthor{Lee2021}'s exact candidate database is not public. We therefore tested these 37 against a rebuilt pool of about 8.9~million structures drawn from the same source databases. Only 13 are present: 11 of 24 closed-shell species, 2 of 11 radicals and none of the 2 ions. For most of the missing species, the limitation lies in the source databases rather than in the ranking, and it falls hardest on radicals and ions that dominate TMC-1 chemistry.

Finally, all features derive from one pipeline~\citep{Lee2021}: a single embedding and a single regression model family, applied to a single source. The two post hoc feature regimes show that the result does not depend on the specific 2021 fit, but not that it would transfer to other embeddings or to other sources.

\subsection{Implications} \label{subsec:implications}
For teams that already maintain a candidate list with model-derived features, the combined ranking offers a better ordering at a negligible cost. It needs no new embedding, no retraining of the abundance model, and only a refit of a seven-feature logistic regression whenever a detection is reported. The feedback-search result (V3) indicates that this refit loop, not any single fixed criterion, is the useful unit.

The results also argue against adopting the textbook active-learning acquisition rule unexamined. In this problem, the standard notion of ``informative" candidates points away from detections, and the useful signal lies in the opposite direction.

Better ranking cannot recover species the list does not contain. The coverage gap for radicals and ions suggests that broadening candidate generation toward astrochemistry-specific sources would yield more than further refinement of the ranking. Constructing such a list lies outside of the scope of a retrospective test: any candidate set assembled now would reflect knowledge of the detections it is evaluated against, and so could only be assessed prospectively. Spectroscopic feasibility, as examined by \citet{ToruShay2025a}, is a complementary constraint that a combined ranking could incorporate directly as an additional feature.

We provide the current ranked list of 100 recommended targets in Table~\ref{tab:targets}, together with the code and frozen inputs needed to reproduce every result~(Section~\ref{sec:methods}). We intend to score this list against detections toward TMC-1 as further survey results are published, and the released materials allow anyone to repeat that test independently.

\section{Conclusions} \label{sec:conc}
We tested, retrospectively and under a pre-registered protocol, whether an active-learning ranking of the published \texttt{umda} recommendation list~\citep{Lee2021} would have anticipated the species detected toward TMC-1 between 2022 January and 2026 September. Our main findings are as follows.
\begin{enumerate}
	\item A combined ranking, a logistic regression on seven released features refit at each origin from the detections known at that time, orders the list better than any single criterion. Its pooled AUC is 0.867, against 0.706 for the strongest baseline (difference 0.160, 95\% interval 0.095 to 0.230). The result holds under a 60-day shift of the date boundary and in two post hoc feature regimes, in which the combined ranking changes by less than 0.01 while the strongest baseline changes in both value and identity.
	\item The advantage is not established at a fixed budget of 100 targets. There the combined ranking recovers 12 later detections and detectability recovers 7, both well above the random expectation of 2.1, but the interval on the difference includes zero.
	\item A searcher that refits on its own observations reaches half of the 63 detections in a median of 200 observations. The best fixed ordering needs 332, and random search would need 717.5 in expectation. All 200 seeds outperform the best fixed ordering.
	\item Uncertainty sampling, the standard active-learning acquisition rule, performs below random (AUC 0.210). Species detected after 2021 lie at low predictive uncertainty, close to known chemistry and at low molecular mass. The combined model learns to weight uncertainty negatively from the detections available at each origin. For target selection aimed at detection, predictive uncertainty is informative in the opposite direction from the one the standard rule assumes.
	\item Of the 101 species in the validation set, 37 do not appear in the recommendation list, and only 13 of those are present in a rebuilt pool drawn from the same source databases. None of the 2 ions and only 2 of the 11 radicals are present. Candidate generation, not ranking, is the binding constraint for the radicals and ions that dominate TMC-1 chemistry.
\end{enumerate}
We provide the current list of 100 recommended targets~(Table~\ref{tab:targets}), together with the code and frozen inputs needed to reproduce every result, and we intend to score that list against new detections as survey results are published.

\begin{acknowledgments}
	\software{
		Python 3.11 (analysis) and 3.10~\citep[embedding and umda reproduction; ][]{Rossum2010},
		NumPy 2.4.4 / 1.26.4~\citep{Harris2020},
		SciPy 1.17.1 / 1.15.3~\citep{Virtanen2020},
		pandas 3.0.2 / 2.3.3~\citep{McKinney2010, pandas2026},
		scikit-learn 1.8.0 / 1.0.2~\citep{Pedregosa2011},
		Matplotlib 3.10.9~\citep{Hunter2007},
		Astropy 8.0.1~\citep{Robitaille2013, PriceWhelan2018, Collaboration2022},
		RDKit 2026.3.6, including its InChI interface~\citep{GregLandrum2026, Heller2015},
		mol2vec 0.1, umda fork at commit 850d944~\citep{Jaeger2018},
		gensim 4.4.0~\citep{Rehurek2010},
		joblib~\citep{joblib2026},
		umda 0.1.0 (\url{https://github.com/laserkelvin/umda})
	}
	
	This research has made use of NASA's Astrophysics Data System Bibliographic Services. We thank \citeauthor{Lee2021} for making the \texttt{umda} code and their recommendation list~\citep{Lee2021, Lee2021a} publicly available; this work would not have been possible without them. The candidate-coverage analysis used structures from QM9~\citep{Ramakrishnan2014}, ZINC15~\citep{Sterling2015}, PubChem~\citep{Kim2020}, the PubChem BioAssay collection as distributed with MoleculeNet~\citep{Wang2011, Wu2018}, the KInetic Database for Astrochemistry~\citep{Wakelam2015}, and the NASA Ames PAH IR Spectroscopic Database~\citep{Boersma2014, Bauschlicher2018, Mattioda2020, Ricca2025}.
	
	The authors declare the use of generative AI in the research and writing process. According to the GAIDeT taxonomy~\citep{Suchikova2025}, the following tasks were delegated to GAI tools uncer full human supervision: literature search, code generation, data curation and organization, reproducibility testing, drafting part of the manuscript text, and proofreading and editing. All decisions were set or approved by the authors. The GAI tool used was Claude~(Anthropic; Opus~5.5 and Sonnet~5.5). Responsibility for the final manuscript lies entirely with the authors.
\end{acknowledgments}

\bibliography{refs}{}

@Article{Lee2021,
  author    = {Lee, Kin Long Kelvin and Patterson, Jacqueline and Burkhardt, Andrew M. and Vankayalapati, Vivek and McCarthy, Michael C. and McGuire, Brett A.},
  journal   = {ApJL},
  title     = {Machine Learning of Interstellar Chemical Inventories},
  year      = {2021},
  issn      = {2041-8213},
  month     = Aug,
  number    = {1},
  pages     = {L6},
  volume    = {917},
  doi       = {10.3847/2041-8213/ac194b},
  publisher = {American Astronomical Society},
}

@Article{McGuire2020,
  author    = {McGuire, Brett A. and Burkhardt, Andrew M. and Loomis, Ryan A. and Shingledecker, Christopher N. and Kelvin Lee, Kin Long and Charnley, Steven B. and Cordiner, Martin A. and Herbst, Eric and Kalenskii, Sergei and Momjian, Emmanuel and Willis, Eric R. and Xue, Ci and Remijan, Anthony J. and McCarthy, Michael C.},
  journal   = {\apjl},
  title     = {Early Science from GOTHAM: Project Overview, Methods, and the Detection of Interstellar Propargyl Cyanide (HCCCH$_2$CN) in TMC-1},
  year      = {2020},
  issn      = {2041-8213},
  month     = Sept,
  number    = {1},
  pages     = {L10},
  volume    = {900},
  doi       = {10.3847/2041-8213/aba632},
  publisher = {American Astronomical Society},
}

@Article{Cernicharo2022,
  author    = {Cernicharo, J. and Agúndez, M. and Cabezas, C. and Fuentetaja, R. and Tercero, B. and Marcelino, N. and Endo, Y. and Pardo, J. R. and de Vicente, P.},
  journal   = {A\&A},
  title     = {Discovery of C$_5$H$^+$ and detection of C$_3$H$^+$ in TMC-1 with the QUIJOTE line survey},
  year      = {2022},
  issn      = {1432-0746},
  month     = Jan,
  pages     = {L16},
  volume    = {657},
  doi       = {10.1051/0004-6361/202142992},
  publisher = {EDP Sciences},
}

@Article{Silva2023,
  author    = {Silva, W. G. D. P. and Cernicharo, J. and Schlemmer, S. and Marcelino, N. and Loison, J.-C. and Agúndez, M. and Gupta, D. and Wakelam, V. and Thorwirth, S. and Cabezas, C. and Tercero, B. and Doménech, J. L. and Fuentetaja, R. and Kim, W.-J. and de Vicente, P. and Asvany, O.},
  journal   = {\aap},
  title     = {Discovery of H$_2$CCCH$^+$ in TMC-1},
  year      = {2023},
  issn      = {1432-0746},
  month     = July,
  pages     = {L1},
  volume    = {676},
  doi       = {10.1051/0004-6361/202347174},
  publisher = {EDP Sciences},
}

@Article{Remijan2023,
  author    = {Remijan, Anthony and Scolati, Haley N. and Burkhardt, Andrew M. and Changala, P. Bryan and Charnley, Steven B. and Cooke, Ilsa R. and Cordiner, Martin A. and Gupta, Harshal and Herbst, Eric and Kelvin Lee, Kin Long and Loomis, Ryan A. and Shingledecker, Christopher N. and Siebert, Mark A. and Xue, Ci and McCarthy, Michael C. and McGuire, Brett A.},
  journal   = {\apjl},
  title     = {Astronomical Detection of the Interstellar Anion C$_{10}$H$^-$ toward TMC-1 from the GOTHAM Large Program on the Green Bank Telescope},
  year      = {2023},
  issn      = {2041-8213},
  month     = Feb,
  number    = {2},
  pages     = {L45},
  volume    = {944},
  doi       = {10.3847/2041-8213/acb648},
  publisher = {American Astronomical Society},
}

@Article{Wenzel2025,
  author    = {Wenzel, Gabi and Gong, Siyuan and Xue, Ci and Changala, P. Bryan and Holdren, Martin S. and Speak, Thomas H. and Stewart, D. Archie and Fried, Zachary T. P. and Willis, Reace H. J. and Bergin, Edwin A. and Burkhardt, Andrew M. and Byrne, Alex N. and Charnley, Steven B. and Lipnicky, Andrew and Loomis, Ryan A. and Shingledecker, Christopher N. and Cooke, Ilsa R. and McCarthy, Michael C. and Remijan, Anthony J. and Wendlandt, Alison E. and McGuire, Brett A.},
  journal   = {\apjl},
  title     = {Discovery of the Seven-ring Polycyclic Aromatic Hydrocarbon Cyanocoronene (C$_{24}$H$_{11}$CN) in GOTHAM Observations of TMC-1},
  year      = {2025},
  issn      = {2041-8213},
  month     = Apr,
  number    = {1},
  pages     = {L36},
  volume    = {984},
  doi       = {10.3847/2041-8213/adc911},
  publisher = {American Astronomical Society},
}

@Article{Jaeger2018,
  author    = {Jaeger, Sabrina and Fulle, Simone and Turk, Samo},
  journal   = {J. Chem. Inf. Model},
  title     = {Mol2vec: Unsupervised Machine Learning Approach with Chemical Intuition},
  year      = {2018},
  issn      = {1549-960X},
  month     = Jan,
  number    = {1},
  pages     = {27--35},
  volume    = {58},
  doi       = {10.1021/acs.jcim.7b00616},
  publisher = {American Chemical Society (ACS)},
}

@Article{Fried2023,
  author    = {Fried, Zachary T. P. and Lee, Kin Long Kelvin and Byrne, Alex N. and McGuire, Brett A.},
  journal   = {Digit. Discov.},
  title     = {Implementation of rare isotopologues into machine learning of the chemical inventory of the solar-type protostellar source IRAS 16293-2422},
  year      = {2023},
  issn      = {2635-098X},
  number    = {4},
  pages     = {952--966},
  volume    = {2},
  doi       = {10.1039/d3dd00020f},
  publisher = {Royal Society of Chemistry (RSC)},
}

@Article{Scolati2023,
  author    = {Scolati, Haley N. and Remijan, Anthony J. and Herbst, Eric and McGuire, Brett A. and Lee, Kin Long Kelvin},
  journal   = {\apj},
  title     = {Explaining the Chemical Inventory of Orion KL through Machine Learning},
  year      = {2023},
  issn      = {1538-4357},
  month     = Dec,
  number    = {2},
  pages     = {108},
  volume    = {959},
  doi       = {10.3847/1538-4357/ad004c},
  publisher = {American Astronomical Society},
}

@Article{ToruShay2025,
  author    = {Toru Shay, Hannah and Scolati, Haley N. and Wenzel, Gabi and Lee, Kin Long Kelvin and Marimuthu, Aravindh N. and McGuire, Brett A.},
  journal   = {\apj},
  title     = {Exploring Effects of Modified Machine Learning Pipelines of Astrochemical Inventories},
  year      = {2025},
  issn      = {1538-4357},
  month     = May,
  number    = {1},
  pages     = {123},
  volume    = {985},
  doi       = {10.3847/1538-4357/adc80b},
  publisher = {American Astronomical Society},
}

@Article{ToruShay2025a,
  author    = {Toru Shay, Hannah and Wenzel, Gabi and Xue, Ci and McGuire, Brett A.},
  journal   = {\apj},
  title     = {Bridging Machine Learning and Spectroscopy: A New Analysis for Astrochemical Target Selection},
  year      = {2025},
  issn      = {1538-4357},
  month     = Oct,
  number    = {1},
  pages     = {75},
  volume    = {993},
  doi       = {10.3847/1538-4357/ae03ae},
  publisher = {American Astronomical Society},
}

@Article{Fried2026,
  author        = {Fried, Zachary T. P. and Loomis, Ryan A. and Jørgensen, Jes K. and Lipnicky, Andrew and Xue, Ci and Wenzel, Gabi and Speak, Thomas H. and McCarthy, Michael C. and McGuire, Brett A.},
  title         = {Automated Assignment and Prediction of Molecules in Astronomical Line Surveys Using Machine-Learning-Based Chemical Embeddings},
  year          = {2026},
  month         = aug,
  archiveprefix = {arXiv},
  copyright     = {arXiv.org perpetual, non-exclusive license},
  doi           = {10.48550/ARXIV.2608.18221},
  eprint        = {2608.18221},
  primaryclass  = {astro-ph.GA},
  publisher     = {arXiv},
}

@Article{Li2026,
  author    = {Li, Guangping and Ou, Chao and Wang, Junzhi and Zhang, Yong and Wang, Zhao},
  journal   = {\aap},
  title     = {From chemical space to observational priority: Predicting detectable molecules in IRC+10216},
  year      = {2026},
  issn      = {1432-0746},
  month     = Mar,
  pages     = {A72},
  volume    = {707},
  doi       = {10.1051/0004-6361/202558794},
  publisher = {EDP Sciences},
}

@Misc{Lee2021a,
  author    = {Lee, Kin Long Kelvin},
  title     = {Recommendations for molecules of interest in the Taurus Molecular Cloud-1},
  year      = {2021},
  copyright = {Creative Commons Attribution 4.0 International},
  doi       = {10.5281/ZENODO.5146276},
  publisher = {Zenodo},
}

@Article{Pedregosa2011,
  author  = {Fabian Pedregosa and Ga{{\"e}}l Varoquaux and Alexandre Gramfort and Vincent Michel and Bertrand Thirion and Olivier Grisel and Mathieu Blondel and Peter Prettenhofer and Ron Weiss and Vincent Dubourg and Jake Vanderplas and Alexandre Passos and David Cournapeau and Matthieu Brucher and Matthieu Perrot and {{\'E}}douard Duchesnay},
  journal = {J. Mach. Learn. Res.},
  title   = {Scikit-learn: Machine Learning in Python},
  year    = {2011},
  number  = {85},
  pages   = {2825--2830},
  volume  = {12},
  url     = {http://jmlr.org/papers/v12/pedregosa11a.html},
}

@Article{McGuire2022,
  author    = {McGuire, Brett A.},
  journal   = {\apjs},
  title     = {2021 Census of Interstellar, Circumstellar, Extragalactic, Protoplanetary Disk, and Exoplanetary Molecules},
  year      = {2022},
  issn      = {1538-4365},
  month     = Mar,
  number    = {2},
  pages     = {30},
  volume    = {259},
  doi       = {10.3847/1538-4365/ac2a48},
  publisher = {American Astronomical Society},
}

@Article{Kaifu2004,
  author    = {Kaifu, Norio and Ohishi, Masatoshi and Kawaguchi, Kentarou and Saito, Shuji and Yamamoto, Satoshi and Miyaji, Takeshi and Miyazawa, Keisuke and Ishikawa, Shin-ichi and Noumaru, Chiaki and Harasawa, Sumiko and Okuda, Michiko and Suzuki, Hiroko},
  journal   = {PASJ},
  title     = {A 8.8–50 GHz Complete Spectral Line Survey toward TMC-1 I. Survey Data},
  year      = {2004},
  issn      = {0004-6264},
  month     = Feb,
  number    = {1},
  pages     = {69--173},
  volume    = {56},
  doi       = {10.1093/pasj/56.1.69},
  publisher = {Oxford University Press (OUP)},
}

@Article{Agundez2013,
  author    = {Agúndez, Marcelino and Wakelam, Valentine},
  journal   = {Chem. Rev.},
  title     = {Chemistry of Dark Clouds: Databases, Networks, and Models},
  year      = {2013},
  issn      = {1520-6890},
  month     = Oct,
  number    = {12},
  pages     = {8710--8737},
  volume    = {113},
  doi       = {10.1021/cr4001176},
  publisher = {American Chemical Society (ACS)},
}

@Article{Agundez2026,
  author        = {Ag\'{u}ndez, Marcelino and Cernicharo, Jose},
  title         = {TMC-1: probing the onset of chemical complexity in space},
  year          = {2026},
  month         = apr,
  archiveprefix = {arXiv},
  copyright     = {Creative Commons Attribution 4.0 International},
  doi           = {10.48550/ARXIV.2604.26439},
  eprint        = {2604.26439},
  primaryclass  = {astro-ph.GA},
  publisher     = {arXiv},
}

@Article{Wootten1980,
  author    = {Wootten, A. and Bozyan, E. P. and Garrett, D. B. and Loren, R. B. and Snell, R. L.},
  journal   = {\apj},
  title     = {Detection of C2H in cold dark clouds},
  year      = {1980},
  issn      = {1538-4357},
  month     = Aug,
  pages     = {844},
  volume    = {239},
  doi       = {10.1086/158168},
  publisher = {American Astronomical Society},
}

@Article{Minh1989,
  author    = {Minh, Y. C. and Irvine, W. M. and Ziurys, L. M.},
  journal   = {\apj},
  title     = {Detection of interstellar hydrogen sulfide in cold, dark clouds},
  year      = {1989},
  issn      = {1538-4357},
  month     = Oct,
  pages     = {L63},
  volume    = {345},
  doi       = {10.1086/185553},
  publisher = {American Astronomical Society},
}

@Article{Schilke1991,
  author   = {{Schilke}, P. and {Walmsley}, C.~M. and {Millar}, T.~J. and {Henkel}, C.},
  journal  = {\aap},
  title    = {{Protonated HCN in molecular clouds.}},
  year     = {1991},
  month    = jul,
  pages    = {487},
  volume   = {247},
  adsurl   = {https://ui.adsabs.harvard.edu/abs/1991A&A...247..487S},
}

@Article{McGonagle1994,
  author    = {McGonagle, Douglas and Irvine, William M. and Ohishi, Masatoshi},
  journal   = {ApJ},
  title     = {Nitrogen sulfide in quiescent dark clouds},
  year      = {1994},
  issn      = {1538-4357},
  month     = Feb,
  pages     = {621},
  volume    = {422},
  doi       = {10.1086/173755},
  publisher = {American Astronomical Society},
}

@Article{Pratap1997,
  author    = {Pratap, P. and Dickens, J. E. and Snell, R. L. and Miralles, M. P. and Bergin, E. A. and Irvine, W. M. and Schloerb, F. P.},
  journal   = {\apj},
  title     = {A Study of the Physics and Chemistry of TMC-1},
  year      = {1997},
  issn      = {1538-4357},
  month     = Sept,
  number    = {2},
  pages     = {862--885},
  volume    = {486},
  doi       = {10.1086/304553},
  publisher = {American Astronomical Society},
}

@Article{Gerin1993,
  author   = {{Gerin}, M. and {Viala}, Y. and {Casoli}, F.},
  journal  = {\aap},
  title    = {{The abundance of nitric oxide in TMC 1.}},
  year     = {1993},
  month    = feb,
  pages    = {212-214},
  volume   = {268},
  adsurl   = {https://ui.adsabs.harvard.edu/abs/1993A&A...268..212G},
}

@Article{Mueller2005,
  author    = {Müller, Holger S.P. and Schlöder, Frank and Stutzki, Jürgen and Winnewisser, Gisbert},
  journal   = {J. Mol. Struct.},
  title     = {The Cologne Database for Molecular Spectroscopy, CDMS: a useful tool for astronomers and spectroscopists},
  year      = {2005},
  issn      = {0022-2860},
  month     = May,
  number    = {1-3},
  pages     = {215--227},
  volume    = {742},
  doi       = {10.1016/j.molstruc.2005.01.027},
  publisher = {Elsevier BV},
}

@Article{Pickett1998,
  author    = {Pickett, H.M. and Poynter, R.L. and Cohen, E.A. and Delitsky, M.L. and Pearson, J.C. and M\"{u}ller, H.S.P.},
  journal   = {J. Quant. Spectrosc. Radiat. Transf.},
  title     = {SUBMILLIMETER, MILLIMETER, AND MICROWAVE SPECTRAL LINE CATALOG},
  year      = {1998},
  issn      = {0022-4073},
  month     = Nov,
  number    = {5},
  pages     = {883--890},
  volume    = {60},
  doi       = {10.1016/s0022-4073(98)00091-0},
  publisher = {Elsevier BV},
}

@Article{Lovas2006,
  author    = {Lovas, F. J. and Hollis, J. M. and Remijan, Anthony J. and Jewell, P. R.},
  journal   = {ApJ},
  title     = {Detection of Ketenimine (CH$_2$CNH) in Sagittarius B2(N) Hot Cores},
  year      = {2006},
  issn      = {1538-4357},
  month     = June,
  number    = {2},
  pages     = {L137--L140},
  volume    = {645},
  doi       = {10.1086/506324},
  publisher = {American Astronomical Society},
}

@Article{Ramakrishnan2014,
  author    = {Ramakrishnan, Raghunathan and Dral, Pavlo O. and Rupp, Matthias and von Lilienfeld, O. Anatole},
  journal   = {Sci. Data},
  title     = {Quantum chemistry structures and properties of 134 kilo molecules},
  year      = {2014},
  issn      = {2052-4463},
  month     = Aug,
  number    = {1},
  volume    = {1},
  doi       = {10.1038/sdata.2014.22},
  publisher = {Springer Science and Business Media LLC},
}

@Article{Sterling2015,
  author    = {Sterling, Teague and Irwin, John J.},
  journal   = {J. Chem. Inf. Model.},
  title     = {ZINC 15 – Ligand Discovery for Everyone},
  year      = {2015},
  issn      = {1549-960X},
  month     = Nov,
  number    = {11},
  pages     = {2324--2337},
  volume    = {55},
  doi       = {10.1021/acs.jcim.5b00559},
  publisher = {American Chemical Society (ACS)},
}

@Article{Kim2020,
  author    = {Kim, Sunghwan and Chen, Jie and Cheng, Tiejun and Gindulyte, Asta and He, Jia and He, Siqian and Li, Qingliang and Shoemaker, Benjamin A and Thiessen, Paul A and Yu, Bo and Zaslavsky, Leonid and Zhang, Jian and Bolton, Evan E},
  journal   = {NAR},
  title     = {PubChem in 2021: new data content and improved web interfaces},
  year      = {2021},
  issn      = {1362-4962},
  month     = Nov,
  number    = {D1},
  pages     = {D1388--D1395},
  volume    = {49},
  doi       = {10.1093/nar/gkaa971},
  publisher = {Oxford University Press (OUP)},
}

@Article{Wang2011,
  author    = {Wang, Y. and Xiao, J. and Suzek, T. O. and Zhang, J. and Wang, J. and Zhou, Z. and Han, L. and Karapetyan, K. and Dracheva, S. and Shoemaker, B. A. and Bolton, E. and Gindulyte, A. and Bryant, S. H.},
  journal   = {NAR},
  title     = {PubChem’s BioAssay Database},
  year      = {2012},
  issn      = {1362-4962},
  month     = Dec,
  number    = {D1},
  pages     = {D400--D412},
  volume    = {40},
  doi       = {10.1093/nar/gkr1132},
  publisher = {Oxford University Press (OUP)},
}

@Article{Wu2018,
  author    = {Wu, Zhenqin and Ramsundar, Bharath and Feinberg, Evan N. and Gomes, Joseph and Geniesse, Caleb and Pappu, Aneesh S. and Leswing, Karl and Pande, Vijay},
  journal   = {Chem. Sci.},
  title     = {MoleculeNet: a benchmark for molecular machine learning},
  year      = {2018},
  issn      = {2041-6539},
  number    = {2},
  pages     = {513--530},
  volume    = {9},
  doi       = {10.1039/c7sc02664a},
  publisher = {Royal Society of Chemistry (RSC)},
}

@Article{Wakelam2015,
  author    = {Wakelam, V. and Loison, J.-C. and Herbst, E. and Pavone, B. and Bergeat, A. and Béroff, K. and Chabot, M. and Faure, A. and Galli, D. and Geppert, W. D. and Gerlich, D. and Gratier, P. and Harada, N. and Hickson, K. M. and Honvault, P. and Klippenstein, S. J. and Picard, S. D. Le and Nyman, G. and Ruaud, M. and Schlemmer, S. and Sims, I. R. and Talbi, D. and Tennyson, J. and Wester, R.},
  journal   = {\apjs},
  title     = {THE 2014 KIDA NETWORK FOR INTERSTELLAR CHEMISTRY},
  year      = {2015},
  issn      = {1538-4365},
  month     = Mar,
  number    = {2},
  pages     = {20},
  volume    = {217},
  doi       = {10.1088/0067-0049/217/2/20},
  publisher = {American Astronomical Society},
}

@Article{Boersma2014,
  author    = {Boersma, C. and Bauschlicher, C. W. and Ricca, A. and Mattioda, A. L. and Cami, J. and Peeters, E. and de Armas, F. Sánchez and Saborido, G. Puerta and Hudgins, D. M. and Allamandola, L. J.},
  journal   = {ApJS},
  title     = {THE NASA AMES PAH IR SPECTROSCOPIC DATABASE VERSION 2.00: UPDATED CONTENT, WEB SITE, AND ON(OFF)LINE TOOLS},
  year      = {2014},
  issn      = {1538-4365},
  month     = Feb,
  number    = {1},
  pages     = {8},
  volume    = {211},
  doi       = {10.1088/0067-0049/211/1/8},
  publisher = {American Astronomical Society},
}

@Article{Bauschlicher2018,
  author    = {Bauschlicher, Charles W. and Ricca, A. and Boersma, C. and Allamandola, L. J.},
  journal   = {ApJS},
  title     = {The NASA Ames PAH IR Spectroscopic Database: Computational Version 3.00 with Updated Content and the Introduction of Multiple Scaling Factors},
  year      = {2018},
  issn      = {1538-4365},
  month     = Feb,
  number    = {2},
  pages     = {32},
  volume    = {234},
  doi       = {10.3847/1538-4365/aaa019},
  publisher = {American Astronomical Society},
}

@Article{Mattioda2020,
  author    = {Mattioda, A. L. and Hudgins, D. M. and Boersma, C. and Bauschlicher, C. W. and Ricca, A. and Cami, J. and Peeters, E. and de Armas, F. Sánchez and Saborido, G. Puerta and Allamandola, L. J.},
  journal   = {ApJS},
  title     = {The NASA Ames PAH IR Spectroscopic Database: The Laboratory Spectra},
  year      = {2020},
  issn      = {1538-4365},
  month     = Dec,
  number    = {2},
  pages     = {22},
  volume    = {251},
  doi       = {10.3847/1538-4365/abc2c8},
  publisher = {American Astronomical Society},
}

@Article{Ricca2025,
  author    = {Ricca, A. and Boersma, C. and Maragkoudakis, A. and Roser, J. E. and Shannon, Matthew J. and Allamandola, L. J. and Bauschlicher, Charles W.},
  journal   = {\apjs},
  title     = {The NASA Ames PAH IR Spectroscopic Database: Computational Version 4.00, Software Tools, Website, and Documentation},
  year      = {2025},
  issn      = {1538-4365},
  month     = Dec,
  number    = {1},
  pages     = {7},
  volume    = {282},
  doi       = {10.3847/1538-4365/ae1c38},
  publisher = {American Astronomical Society},
}

@Book{Rossum2010,
  author    = {{van Rossum}, Guido and Drake, Fred L.},
  publisher = {CreateSpace},
  title     = {Python 3 Reference Manual},
  year      = {2009},
  address   = {Scotts Valley, CA},
}

@Article{Harris2020,
  author    = {Harris, Charles R. and Millman, K. Jarrod and van der Walt, Stéfan J. and Gommers, Ralf and Virtanen, Pauli and Cournapeau, David and Wieser, Eric and Taylor, Julian and Berg, Sebastian and Smith, Nathaniel J. and Kern, Robert and Picus, Matti and Hoyer, Stephan and van Kerkwijk, Marten H. and Brett, Matthew and Haldane, Allan and del Río, Jaime Fernández and Wiebe, Mark and Peterson, Pearu and Gérard-Marchant, Pierre and Sheppard, Kevin and Reddy, Tyler and Weckesser, Warren and Abbasi, Hameer and Gohlke, Christoph and Oliphant, Travis E.},
  journal   = {Nature},
  title     = {Array programming with NumPy},
  year      = {2020},
  issn      = {1476-4687},
  month     = Sept,
  number    = {7825},
  pages     = {357--362},
  volume    = {585},
  doi       = {10.1038/s41586-020-2649-2},
  publisher = {Springer Science and Business Media LLC},
}

@Article{Virtanen2020,
  author    = {Virtanen, Pauli and Gommers, Ralf and Oliphant, Travis E. and Haberland, Matt and Reddy, Tyler and Cournapeau, David and Burovski, Evgeni and Peterson, Pearu and Weckesser, Warren and Bright, Jonathan and van der Walt, Stéfan J. and Brett, Matthew and Wilson, Joshua and Millman, K. Jarrod and Mayorov, Nikolay and Nelson, Andrew R. J. and Jones, Eric and Kern, Robert and Larson, Eric and Carey, C J and Polat, İlhan and Feng, Yu and Moore, Eric W. and VanderPlas, Jake and Laxalde, Denis and Perktold, Josef and Cimrman, Robert and Henriksen, Ian and Quintero, E. A. and Harris, Charles R. and Archibald, Anne M. and Ribeiro, Antônio H. and Pedregosa, Fabian and van Mulbregt, Paul and Vijaykumar, Aditya and Bardelli, Alessandro Pietro and Rothberg, Alex and Hilboll, Andreas and Kloeckner, Andreas and Scopatz, Anthony and Lee, Antony and Rokem, Ariel and Woods, C. Nathan and Fulton, Chad and Masson, Charles and Häggström, Christian and Fitzgerald, Clark and Nicholson, David A. and Hagen, David R. and Pasechnik, Dmitrii V. and Olivetti, Emanuele and Martin, Eric and Wieser, Eric and Silva, Fabrice and Lenders, Felix and Wilhelm, Florian and Young, G. and Price, Gavin A. and Ingold, Gert-Ludwig and Allen, Gregory E. and Lee, Gregory R. and Audren, Hervé and Probst, Irvin and Dietrich, Jörg P. and Silterra, Jacob and Webber, James T and Slavič, Janko and Nothman, Joel and Buchner, Johannes and Kulick, Johannes and Schönberger, Johannes L. and de Miranda Cardoso, José Vinícius and Reimer, Joscha and Harrington, Joseph and Rodríguez, Juan Luis Cano and Nunez-Iglesias, Juan and Kuczynski, Justin and Tritz, Kevin and Thoma, Martin and Newville, Matthew and Kümmerer, Matthias and Bolingbroke, Maximilian and Tartre, Michael and Pak, Mikhail and Smith, Nathaniel J. and Nowaczyk, Nikolai and Shebanov, Nikolay and Pavlyk, Oleksandr and Brodtkorb, Per A. and Lee, Perry and McGibbon, Robert T. and Feldbauer, Roman and Lewis, Sam and Tygier, Sam and Sievert, Scott and Vigna, Sebastiano and Peterson, Stefan and More, Surhud and Pudlik, Tadeusz and Oshima, Takuya and Pingel, Thomas J. and Robitaille, Thomas P. and Spura, Thomas and Jones, Thouis R. and Cera, Tim and Leslie, Tim and Zito, Tiziano and Krauss, Tom and Upadhyay, Utkarsh and Halchenko, Yaroslav O. and Vázquez-Baeza, Yoshiki},
  journal   = {Nat. Methods},
  title     = {SciPy 1.0: fundamental algorithms for scientific computing in Python},
  year      = {2020},
  issn      = {1548-7105},
  month     = Feb,
  number    = {3},
  pages     = {261--272},
  volume    = {17},
  doi       = {10.1038/s41592-019-0686-2},
  publisher = {Springer Science and Business Media LLC},
}

@InProceedings{McKinney2010,
  author     = {McKinney, Wes},
  booktitle  = {Proceedings of the 9th Python in Science Conference},
  title      = {Data Structures for Statistical Computing in Python},
  year       = {2010},
  pages      = {56--61},
  publisher  = {SciPy},
  series     = {SciPy},
  collection = {SciPy},
  doi        = {10.25080/majora-92bf1922-00a},
  issn       = {2575-9752},
}

@Misc{pandas2026,
  author    = {{The pandas development team}},
  title     = {pandas-dev/pandas: Pandas},
  year      = {2026},
  copyright = {BSD 3-Clause "New" or "Revised" License},
  doi       = {10.5281/ZENODO.3509134},
  publisher = {Zenodo},
}

@Article{Hunter2007,
  author    = {Hunter, John D.},
  journal   = {Comput. Sci. Eng.},
  title     = {Matplotlib: A 2D Graphics Environment},
  year      = {2007},
  issn      = {1521-9615},
  number    = {3},
  pages     = {90--95},
  volume    = {9},
  doi       = {10.1109/mcse.2007.55},
  publisher = {Institute of Electrical and Electronics Engineers (IEEE)},
}

@Article{Robitaille2013,
  author    = {{Astropy Collaboration} and Robitaille, Thomas P. and Tollerud, Erik J. and Greenfield, Perry and Droettboom, Michael and Bray, Erik and Aldcroft, Tom and Davis, Matt and Ginsburg, Adam and Price-Whelan, Adrian M. and Kerzendorf, Wolfgang E. and Conley, Alexander and Crighton, Neil and Barbary, Kyle and Muna, Demitri and Ferguson, Henry and Grollier, Frédéric and Parikh, Madhura M. and Nair, Prasanth H. and Günther, Hans M. and Deil, Christoph and Woillez, Julien and Conseil, Simon and Kramer, Roban and Turner, James E. H. and Singer, Leo and Fox, Ryan and Weaver, Benjamin A. and Zabalza, Victor and Edwards, Zachary I. and Azalee Bostroem, K. and Burke, D. J. and Casey, Andrew R. and Crawford, Steven M. and Dencheva, Nadia and Ely, Justin and Jenness, Tim and Labrie, Kathleen and Lim, Pey Lian and Pierfederici, Francesco and Pontzen, Andrew and Ptak, Andy and Refsdal, Brian and Servillat, Mathieu and Streicher, Ole},
  journal   = {\aap},
  title     = {Astropy: A community Python package for astronomy},
  year      = {2013},
  issn      = {1432-0746},
  month     = Sept,
  pages     = {A33},
  volume    = {558},
  doi       = {10.1051/0004-6361/201322068},
  publisher = {EDP Sciences},
}

@Article{PriceWhelan2018,
  author    = {{Astropy Collaboration} and Price-Whelan, A. M. and Sipőcz, B. M. and Günther, H. M. and Lim, P. L. and Crawford, S. M. and Conseil, S. and Shupe, D. L. and Craig, M. W. and Dencheva, N. and Ginsburg, A. and VanderPlas, J. T. and Bradley, L. D. and Pérez-Suárez, D. and de Val-Borro, M. and Aldcroft, T. L. and Cruz, K. L. and Robitaille, T. P. and Tollerud, E. J. and Ardelean, C. and Babej, T. and Bach, Y. P. and Bachetti, M. and Bakanov, A. V. and Bamford, S. P. and Barentsen, G. and Barmby, P. and Baumbach, A. and Berry, K. L. and Biscani, F. and Boquien, M. and Bostroem, K. A. and Bouma, L. G. and Brammer, G. B. and Bray, E. M. and Breytenbach, H. and Buddelmeijer, H. and Burke, D. J. and Calderone, G. and Rodríguez, J. L. Cano and Cara, M. and Cardoso, J. V. M. and Cheedella, S. and Copin, Y. and Corrales, L. and Crichton, D. and D’Avella, D. and Deil, C. and Depagne, É. and Dietrich, J. P. and Donath, A. and Droettboom, M. and Earl, N. and Erben, T. and Fabbro, S. and Ferreira, L. A. and Finethy, T. and Fox, R. T. and Garrison, L. H. and Gibbons, S. L. J. and Goldstein, D. A. and Gommers, R. and Greco, J. P. and Greenfield, P. and Groener, A. M. and Grollier, F. and Hagen, A. and Hirst, P. and Homeier, D. and Horton, A. J. and Hosseinzadeh, G. and Hu, L. and Hunkeler, J. S. and Ivezić, Ž. and Jain, A. and Jenness, T. and Kanarek, G. and Kendrew, S. and Kern, N. S. and Kerzendorf, W. E. and Khvalko, A. and King, J. and Kirkby, D. and Kulkarni, A. M. and Kumar, A. and Lee, A. and Lenz, D. and Littlefair, S. P. and Ma, Z. and Macleod, D. M. and Mastropietro, M. and McCully, C. and Montagnac, S. and Morris, B. M. and Mueller, M. and Mumford, S. J. and Muna, D. and Murphy, N. A. and Nelson, S. and Nguyen, G. H. and Ninan, J. P. and Nöthe, M. and Ogaz, S. and Oh, S. and Parejko, J. K. and Parley, N. and Pascual, S. and Patil, R. and Patil, A. A. and Plunkett, A. L. and Prochaska, J. X. and Rastogi, T. and Janga, V. Reddy and Sabater, J. and Sakurikar, P. and Seifert, M. and Sherbert, L. E. and Sherwood-Taylor, H. and Shih, A. Y. and Sick, J. and Silbiger, M. T. and Singanamalla, S. and Singer, L. P. and Sladen, P. H. and Sooley, K. A. and Sornarajah, S. and Streicher, O. and Teuben, P. and Thomas, S. W. and Tremblay, G. R. and Turner, J. E. H. and Terrón, V. and Kerkwijk, M. H. van and de la Vega, A. and Watkins, L. L. and Weaver, B. A. and Whitmore, J. B. and Woillez, J. and Zabalza, V. and},
  journal   = {\aj},
  title     = {The Astropy Project: Building an Open-science Project and Status of the v2.0 Core Package},
  year      = {2018},
  issn      = {1538-3881},
  month     = Aug,
  number    = {3},
  pages     = {123},
  volume    = {156},
  doi       = {10.3847/1538-3881/aabc4f},
  publisher = {American Astronomical Society},
}

@Article{Collaboration2022,
  author    = {{Astropy Collaboration} and Price-Whelan, Adrian M. and Lim, Pey Lian and Earl, Nicholas and Starkman, Nathaniel and Bradley, Larry and Shupe, David L. and Patil, Aarya A. and Corrales, Lia and Brasseur, C. E. and Nöthe, Maximilian and Donath, Axel and Tollerud, Erik and Morris, Brett M. and Ginsburg, Adam and Vaher, Eero and Weaver, Benjamin A. and Tocknell, James and Jamieson, William and van Kerkwijk, Marten H. and Robitaille, Thomas P. and Merry, Bruce and Bachetti, Matteo and Günther, H. Moritz and Aldcroft, Thomas L. and Alvarado-Montes, Jaime A. and Archibald, Anne M. and Bódi, Attila and Bapat, Shreyas and Barentsen, Geert and Bazán, Juanjo and Biswas, Manish and Boquien, Médéric and Burke, D. J. and Cara, Daria and Cara, Mihai and Conroy, Kyle E and Conseil, Simon and Craig, Matthew W. and Cross, Robert M. and Cruz, Kelle L. and D’Eugenio, Francesco and Dencheva, Nadia and Devillepoix, Hadrien A. R. and Dietrich, Jörg P. and Eigenbrot, Arthur Davis and Erben, Thomas and Ferreira, Leonardo and Foreman-Mackey, Daniel and Fox, Ryan and Freij, Nabil and Garg, Suyog and Geda, Robel and Glattly, Lauren and Gondhalekar, Yash and Gordon, Karl D. and Grant, David and Greenfield, Perry and Groener, Austen M. and Guest, Steve and Gurovich, Sebastian and Handberg, Rasmus and Hart, Akeem and Hatfield-Dodds, Zac and Homeier, Derek and Hosseinzadeh, Griffin and Jenness, Tim and Jones, Craig K. and Joseph, Prajwel and Kalmbach, J. Bryce and Karamehmetoglu, Emir and Kałuszyński, Mikołaj and Kelley, Michael S. P. and Kern, Nicholas and Kerzendorf, Wolfgang E. and Koch, Eric W. and Kulumani, Shankar and Lee, Antony and Ly, Chun and Ma, Zhiyuan and MacBride, Conor and Maljaars, Jakob M. and Muna, Demitri and Murphy, N. A. and Norman, Henrik and O’Steen, Richard and Oman, Kyle A. and Pacifici, Camilla and Pascual, Sergio and Pascual-Granado, J. and Patil, Rohit R. and Perren, Gabriel I and Pickering, Timothy E. and Rastogi, Tanuj and Roulston, Benjamin R. and Ryan, Daniel F and Rykoff, Eli S. and Sabater, Jose and Sakurikar, Parikshit and Salgado, Jesús and Sanghi, Aniket and Saunders, Nicholas and Savchenko, Volodymyr and Schwardt, Ludwig and Seifert-Eckert, Michael and Shih, Albert Y. and Jain, Anany Shrey and Shukla, Gyanendra and Sick, Jonathan and Simpson, Chris and Singanamalla, Sudheesh and Singer, Leo P. and Singhal, Jaladh and Sinha, Manodeep and Sipőcz, Brigitta M. and Spitler, Lee R. and Stansby, David and Streicher, Ole and Šumak, Jani and Swinbank, John D. and Taranu, Dan S. and Tewary, Nikita and Tremblay, Grant R. and Val-Borro, Miguel de and Van Kooten, Samuel J. and Vasović, Zlatan and Verma, Shresth and de Miranda Cardoso, José Vinícius and Williams, Peter K. G. and Wilson, Tom J. and Winkel, Benjamin and Wood-Vasey, W. M. and Xue, Rui and Yoachim, Peter and Zhang, Chen and Zonca, Andrea and},
  journal   = {ApJ},
  title     = {The Astropy Project: Sustaining and Growing a Community-oriented Open-source Project and the Latest Major Release (v5.0) of the Core Package},
  year      = {2022},
  issn      = {1538-4357},
  month     = Aug,
  number    = {2},
  pages     = {167},
  volume    = {935},
  doi       = {10.3847/1538-4357/ac7c74},
  publisher = {American Astronomical Society},
}

@Misc{GregLandrum2026,
  author       = {Landrum, Greg and Tosco, Paolo and Kelley, Brian and others},
  howpublished = {Zenodo},
  title        = {{RDKit: Open-source cheminformatics}},
  year         = {2026},
  doi          = {10.5281/zenodo.591637},
  version      = {2026.03.6},
}

@Article{Heller2015,
  author    = {Heller, Stephen R and McNaught, Alan and Pletnev, Igor and Stein, Stephen and Tchekhovskoi, Dmitrii},
  journal   = {J. Cheminform.},
  title     = {InChI, the IUPAC International Chemical Identifier},
  year      = {2015},
  issn      = {1758-2946},
  month     = May,
  number    = {1},
  volume    = {7},
  doi       = {10.1186/s13321-015-0068-4},
  publisher = {Springer Science and Business Media LLC},
}

@InProceedings{Rehurek2010,
  author    = {Radim {\v R}eh{\r u}{\v r}ek and Petr Sojka},
  booktitle = {{Proceedings of the LREC 2010 Workshop on New Challenges for NLP Frameworks}},
  title     = {{Software Framework for Topic Modelling with Large Corpora}},
  year      = {2010},
  address   = {Valletta, Malta},
  month     = May,
  note      = {\url{http://is.muni.cz/publication/884893/en}},
  pages     = {45--50},
  publisher = {ELRA},
  day       = {22},
  language  = {English},
}

@Misc{joblib2026,
  author    = {{joblib developers}},
  title     = {joblib},
  year      = {2026},
  copyright = {BSD 3-Clause "New" or "Revised" License},
  doi       = {10.5281/ZENODO.14915601},
  publisher = {Zenodo},
}

@Article{Suchikova2025,
  author    = {Suchikova, Yana and Tsybuliak, Natalia and Teixeira da Silva, Jaime A. and Nazarovets, Serhii},
  journal   = {Account. Res.},
  title     = {GAIDeT (Generative AI Delegation Taxonomy): A taxonomy for humans to delegate tasks to generative artificial intelligence in scientific research and publishing},
  year      = {2025},
  issn      = {1545-5815},
  month     = Aug,
  number    = {3},
  volume    = {33},
  doi       = {10.1080/08989621.2025.2544331},
  publisher = {Informa UK Limited},
}

@Software{zenodo,
  author    = {El-Kholy, R. I. and Hayman, Z. M.},
  doi       = {10.5281/zenodo.23312984},
  month     = oct,
  publisher = {Zenodo},
  title     = {TMC-1: ranking unseen interstellar-molecule candidates with a frozen model (code and data)},
  url       = {https://doi.org/10.5281/zenodo.23312984},
  version   = {v1.0.0,},
  year      = {2026},
}
\bibliographystyle{aasjournalv7.1}
\end{document}